\documentclass[aps,preprint]{revtex4}%
\usepackage{amsfonts}%
\usepackage{amsmath}%
\usepackage{amssymb}%
\usepackage{graphicx}
\providecommand{\U}[1]{\protect\rule{.1in}{.1in}}

\begin{document}
\preprint{ }
\title[ ]{Excitonic coherent states: symmetries and thermalization}
\author{Diego Julio Cirilo-Lombardo}

\affiliation{Bogoliubov Laboratory of Theoretical Physics, Joint Institute for Nuclear
Research, Dubna 141980, Russian Federation}
\author{}
\affiliation{}
\author{}
\affiliation{}

\volumeyear{year}
\volumenumber{number}
\issuenumber{number}
\eid{identifier}
\date[Date text]{date}
\received[Received text]{date}

\revised[Revised text]{date}

\accepted[Accepted text]{date}

\published[Published text]{date}

\startpage{101}
\endpage{102}
\maketitle
\tableofcontents

\section{Introduction}

In previous work [1] new bounded coherent states construction, based on a
Keldysh conjecture, was introduced. As was shown in [1] the particular group
structure arising from the model leads to new symmetry transformations for the
coherent states system. As was shown, the emergent new symmetry transformation
is reminiscent of the Bogoliubov ones and was successfully applied to describe
an excitonic system showing that is intrinsically related to the stability and
its general physical behavior. The group theoretical structure of the model
permits to analyze its thermal properties in theoretical frameworks that arise
as a consequence of the definition of the squeezed-coherent states as
transformed vacua under the automorphism group of the commutation relations,
as the thermofield dynamics case given by Umezawa and other similar
developments [5]. On the other hand, the idea of a possible Bose--Einstein
condensation (BEC) of excitons in semiconductors has attracted the attention
of both experimentalists and theoreticians for more than three decades being
one of the main questions what happens with the influence of non zero
temperature in the case that such condensation really exists [2]. In this
paper we considered the theoretical treatment of the excitonic behaviour by
mean a new coherent state construction of bounded states in a quantum field
theoretical context. The possibility to introduce the coherent states in a
physical system of excitons is mainly fundamented by the idea of "exciton
state splitting" and "exciton wave function" introduced by L.D.Keldysh [3]
earlier and developed by us in [1]. This paper is organized as follows: in
Section I we make basically a short review of our earlier work[1] adding
several new comments and concepts necessary to the clear understanding of this
approach and Section II\ is devoted to introduce the thermalization of
excitonic coherent states by mean an specific unitary transformation described
by a composed displacement operator.

\section{Exciton model: review and new concepts}

\subsection{General description}

As was shown before [1], our starting point is based in the following
splitting of the fermionic state in the material to be considered%

\begin{equation}
\psi_{\alpha}\left(  \mathbf{x}\right)  \equiv\psi_{\alpha}^{\left(  e\right)
}\left(  \mathbf{x}\right)  +\psi_{\alpha}^{\dagger\left(  h\right)  }\left(
\mathbf{x}\right) \tag{1}%
\end{equation}
with $\psi_{\alpha}^{\left(  e\right)  }\left(  \mathbf{x}\right)
\equiv\underset{j>j_{0}}{%
{\displaystyle\sum}
}a_{j}\chi_{\alpha}^{j}\left(  \mathbf{x}\right)  ,\ \ \psi_{\alpha}%
^{\dagger\left(  h\right)  }\left(  \mathbf{x}\right)  \equiv\underset{j\leq
j_{0}}{%
{\displaystyle\sum}
}a_{j}\chi_{\alpha}^{j}\left(  \mathbf{x}\right)  $ where: $\left[
a_{j}^{\dagger},a_{j^{\prime}}\right]  _{+}=\delta_{jj^{\prime}}%
,\ \ \ \ \left[  a_{j},a_{j^{\prime}}\right]  _{+}=0\ \ \ $We have defined
$\chi_{\alpha}^{j}\left(  \mathbf{x}\right)  $ the basic functions of
Hartree-Fock (HF) of the system and the indices $j>j_{0}$ and $j\leq j_{0}$
numerate the bounded states from the electronic zone and the free states respectively.

\begin{quote}
Remark 1: definition (1) describes correctly the excitonic operator being the
same operator acting in the characteristic zones. Then, in sharp contrast with
the traditionally accepted use of different operators for electron and hole
respectively, the construction (1) avoid all type of overcounting and spurious
states that are clearly non physical.
\end{quote}

Let us consider, without lose generality and only to exemplify in concrete
cases, the symmetries of a periodic system (e.g. crystal) the functions of
Hartree-Fock (HF) \ take the form of a Bloch state $\chi_{j\alpha}\left(
\mathbf{x}\right)  =e^{i\mathbf{P\cdot x}}u_{\mathbf{P}l\alpha}$ where $P$ is
the quasi-momentum and $l$ is the number of zone, such that: $j=\left\{
\mathbf{P,}l\right\}  .$ If the case is for a non-metallic crystal, then the
sum in $j\leq j_{0}$ correspond to a sum over all $\mathbf{P}$ which live in
the 1$^{st}$ Brillouin zone. The HF\ functions obey the HF\ equation%
\begin{equation}
\int h_{\alpha\beta}\left(  \mathbf{x,x}^{\prime}\right)  \chi_{j}^{\beta
}\left(  \mathbf{x}^{\prime}\right)  d^{3}\mathbf{x}^{\prime}=\varepsilon
_{j}\chi_{j\alpha}\left(  \mathbf{x}\right) \tag{2}%
\end{equation}
with $h_{\alpha\beta}\left(  \mathbf{x,x}^{\prime}\right)  =\delta
_{\alpha\beta}\delta\left(  \mathbf{x,x}^{\prime}\right)  \left\{
\frac{-\hbar^{2}}{2m_{0}}\nabla^{2}-\sum\frac{Z_{k}e^{2}}{\left\vert
\mathbf{R}_{n,k}-\mathbf{x}\right\vert }+\frac{e^{2}}{2}\int\frac
{g_{\beta\beta}\left(  \mathbf{y,y}\right)  d^{3}\mathbf{y}}{\left\vert
\mathbf{x-y}\right\vert }\right\}  -e^{2}\frac{g_{\alpha\beta}\left(
\mathbf{x,x}^{\prime}\right)  }{\left\vert \mathbf{x-x}^{\prime}\right\vert }$
being the HF\ operator, where we define $g_{\alpha\beta}\left(  \mathbf{x,x}%
^{\prime}\right)  \equiv\underset{j\leq j_{0}}{%
{\displaystyle\sum}
}\chi_{j\alpha}\left(  \mathbf{x}\right)  \chi_{\beta}^{j}\left(
\mathbf{x}^{\prime}\right)  $

\begin{quote}
Remark 2: the important observation here (in concordance with our remark about
expression (1)) is that the hamiltonian is not the sum of several terms
involving electrons , holes etc. etc. as separate entities, as is currently
taken in the literature: only the state defined in expression (1) is involved
into the Hamiltonian $h_{\alpha\beta}\left(  \mathbf{x,x}^{\prime}\right)  $.
\end{quote}

\subsection{Exciton wave equation}

Due to the composite characteristic of the excitonic state, firstly we have
particular interest in the 2-particles 2-times Green functions%
\begin{equation}
G_{\alpha\beta,\gamma\delta}^{\left(  2\right)  }\left(  \mathbf{x,y,}%
t;\mathbf{x}^{\prime}\mathbf{,y}^{\prime},t^{\prime}\right)  =-\frac{i}{\hbar
}\left\langle T\psi_{\alpha}^{\dagger}\left(  \mathbf{x,}t\right)  \psi
_{\beta}\left(  \mathbf{y,}t\right)  \psi_{\gamma}^{\dagger}\left(
\mathbf{x}^{\prime},t^{\prime}\right)  \psi_{\delta}\left(  \mathbf{y}%
^{\prime}\mathbf{,}t^{\prime}\right)  \right\rangle _{0}\tag{3}%
\end{equation}
where $\psi_{\beta}\left(  \mathbf{y,}t\right)  $ are Heisenberg operators and
$\left\langle T\text{ ....}\right\rangle _{0}$ chronological product. The
second important point in the CS excitonic formulation is due to
the\ observation pointed out in [2], that the $G_{\alpha\beta,\gamma\delta
}^{\left(  2\right)  }\left(  \mathbf{x,y,}t;\mathbf{x}^{\prime}%
\mathbf{,y}^{\prime},t^{\prime}\right)  $ can be written as%
\begin{equation}
i\hbar G_{\alpha\beta,\gamma\delta}^{\left(  2\right)  }\left(  \mathbf{x,y,}%
t;\mathbf{x}^{\prime}\mathbf{,y}^{\prime},t^{\prime}\right)  =-\underset
{\mathbf{P}J}{%
{\displaystyle\sum}
}\left\{
\begin{array}
[c]{c}%
\varphi_{\alpha\beta}^{J\mathbf{P}}\left(  \mathbf{x,y}\right)  \varphi
_{\gamma\delta}^{J\mathbf{P\star}}\left(  \mathbf{x}^{\prime}\mathbf{,y}%
^{\prime}\right)  e^{\frac{i}{\hbar}\left(  \frac{\mathbf{P}\cdot\left(
\mathbf{x+y-x}^{\prime}-\mathbf{y}^{\prime}\right)  }{2}-E_{J\mathbf{P}%
}(t-t^{\prime})\right)  }\text{ ,\ \ }t>t^{\prime}\\
\varphi_{\alpha\beta}^{J\mathbf{P\star}}\left(  \mathbf{x,y}\right)
\varphi_{\gamma\delta}^{J\mathbf{P}}\left(  \mathbf{x}^{\prime}\mathbf{,y}%
^{\prime}\right)  e^{\frac{-i}{\hbar}\left(  \frac{\mathbf{P}\cdot\left(
\mathbf{x+y-x}^{\prime}-\mathbf{y}^{\prime}\right)  }{2}-E_{J\mathbf{P}%
}(t-t^{\prime})\right)  }\text{,\ \ }t<t^{\prime}%
\end{array}
\right. \tag{4}%
\end{equation}
here is easily seen that%
\begin{equation}
e^{\frac{i}{\hbar}\frac{\mathbf{P}\cdot\left(  \mathbf{x+y}\right)  }{2}%
}\varphi_{\alpha\beta}^{J\mathbf{P}}\left(  \mathbf{x,y}\right)  =\left\langle
0\left\vert \psi_{\alpha}^{\dagger}\left(  \mathbf{x}\right)  \psi_{\beta
}\left(  \mathbf{y}\right)  \right\vert J\mathbf{P}\right\rangle \tag{5}%
\end{equation}
Then, the above expression can be assumed as the basic wave function of the
exciton\footnote{Notice that form eq.(8) the factorization in pairs of the two
times/two field Green's functions is autometically assumed.} Taking account of
the symmetries involved, the Von Karman periodic conditions are $\varphi
_{\alpha\beta}^{J\mathbf{P}}\left(  \mathbf{x+R}_{n}\mathbf{,y+R}_{n}\right)
=\varphi_{\alpha\beta}^{J\mathbf{P}}\left(  \mathbf{x,y}\right)  $ with
$\mathbf{R}_{n}$ a characteristic vector of the crystal lattice. Fourier
transforming (4)\ in the time we obtain%
\begin{equation}
G_{\alpha\beta,\gamma\delta}^{\left(  2\right)  }\left(  \mathbf{x,y;x}%
^{\prime}\mathbf{,y}^{\prime};\mathbf{P}E\right)  =\underset{J,\delta
\rightarrow+0}{%
{\displaystyle\sum}
}\frac{2E_{J\mathbf{P}}}{\underset{}{E^{2}-\left(  E_{J\mathbf{P}}%
-i\delta\right)  ^{2}}}\varphi_{\alpha\beta}^{J\mathbf{P\star}}\left(
\mathbf{x,y}\right)  \varphi_{\gamma\delta}^{J\mathbf{P}}\left(
\mathbf{x}^{\prime}\mathbf{,y}^{\prime}\right) \tag{6}%
\end{equation}
notice that, due the free field form, this formulas are independent on the
specific form of the hamiltonian considered.

\begin{quote}
Remark 3: due the composite and extended character of the exciton systems, the
wave functions defined in (5) have a non-local behaviour in general
\end{quote}

\subsection{Excitonic coherent state construction}

It is well known that CS provide naturally a close connection between
classical and quantum formulations of a given system [4]. As is well known,
the importance of coherent states in physics, and particularly in condensed
matter physics, is huge. All the physical processes where the quantum world is
macroscopically manifested (as in BEC or laser physics) can be faithfully
described by coherent states due the semiclassical behaviour, temporal
stability and other mathematical requisites needed in the quantum field
theoretical framework. There exist three standard definitions in the
construction of coherent states. The most suitable for our proposes here is by
mean a "displacing operator" acting over the vacuum (specific fiducial
vector). The unitary operators%

\begin{align}
B_{J\mathbf{P}}  & =\frac{1}{V}\int\psi^{\alpha\dagger}\left(  \mathbf{x}%
\right)  \varphi_{\alpha\beta}^{J\mathbf{P}}\left(  \mathbf{x,y}\right)
e^{\frac{i}{\hbar}\frac{\mathbf{P}\cdot\left(  \mathbf{x+y}\right)  }{2}}%
\psi^{\beta}\left(  \mathbf{y}\right)  d^{3}\mathbf{x}d^{3}\mathbf{y,}\text{
\ \ }\tag{7}\\
B_{J\mathbf{P}}^{\dagger}  & =\frac{1}{V}\int e^{\frac{-i}{\hbar}%
\frac{\mathbf{P}\cdot\left(  \mathbf{x+y}\right)  }{2}}\psi^{\alpha\dagger
}\left(  \mathbf{x}\right)  \varphi_{\alpha\beta}^{\dagger J\mathbf{P}}\left(
\mathbf{x,y}\right)  \psi^{\beta}\left(  \mathbf{y}\right)  d^{3}%
\mathbf{x}d^{3}\mathbf{y}\nonumber
\end{align}
where $\varphi_{\alpha\beta}^{\dagger J\mathbf{P}}\left(  \mathbf{x,y}\right)
=\left[  \varphi_{\beta\alpha}^{J\mathbf{P}}\left(  \mathbf{y,x}\right)
\right]  ^{\star},\left(  V=\text{ normalized volume}\right)  $and the
commutation relations takes the following form
\begin{gather}
\left[  B_{J\mathbf{P}},B_{J^{\prime}\mathbf{P}^{\prime}}^{\dagger}\right]
=\delta_{JJ^{\prime}}\delta_{\mathbf{PP}^{\prime}}-\tag{8}\\
-\left\{  \frac{1}{V}\int\psi^{\alpha\dagger\left(  e\right)  }\left(
\mathbf{x}\right)  e^{\frac{i}{2\hbar}\mathbf{P}\cdot\mathbf{x}}%
\varphi_{\alpha\gamma}^{J\mathbf{P}}\left(  \mathbf{x,z}\right)  e^{\frac
{-i}{2\hbar}\left(  \mathbf{P-P}^{\prime}\right)  \cdot\mathbf{z}}%
\varphi_{\gamma\beta}^{\dagger J^{\prime}\mathbf{P}^{\prime}}\left(
\mathbf{z,y}\right)  e^{\frac{-i}{2\hbar}\mathbf{P}^{\prime}\cdot\mathbf{y}%
}\psi^{\beta\left(  e\right)  }\left(  \mathbf{y}\right)  +\right. \nonumber\\
\left.  +\frac{1}{V}\int\psi^{\alpha\dagger\left(  h\right)  }\left(
\mathbf{x}\right)  e^{\frac{-i}{2\hbar}\mathbf{P}^{\prime}\cdot\mathbf{y}%
}\varphi_{\gamma\beta}^{\dagger J^{\prime}\mathbf{P}^{\prime}}\left(
\mathbf{y,z}\right)  e^{\frac{-i}{2\hbar}\left(  \mathbf{P-P}^{\prime}\right)
\cdot\mathbf{z}}\varphi_{\alpha\gamma}^{J\mathbf{P}}\left(  \mathbf{z,x}%
\right)  e^{\frac{i}{2\hbar}\mathbf{P}\cdot\mathbf{x}}\psi^{\beta\left(
h\right)  }\left(  \mathbf{y}\right)  \right\}  d^{3}\mathbf{x}d^{3}%
\mathbf{y}d^{3}\mathbf{z}\nonumber
\end{gather}
indicating exactly the intrincated interplay in the electron-hole
system.(notice the lack of canonicity). Although the complexity of expression
(8) we take advantage of the unitarity of the $B_{J\mathbf{P}}$ (7)
constructing the coherent states as%
\begin{equation}
\left\vert \beta,J\mathbf{P}\right\rangle =\exp\left\{  \beta B_{J\mathbf{P}%
}^{\dagger}e^{iE_{J\mathbf{P}}t/\hbar}-\beta^{\star}B_{J\mathbf{P}%
}e^{-iE_{J\mathbf{P}}t/\hbar}\right\}  \left\vert 0\right\rangle
\equiv\left\vert \varphi\right\rangle \tag{9}%
\end{equation}
where, after the use of equation (1), the explicit form of the displacement
operator is as follows%

\begin{align}
D_{\varphi} &  =\exp\left[  \int\psi^{\alpha\dagger\left(  e\right)  }\left(
\mathbf{x}\right)  \varphi_{\alpha\beta}\left(  \mathbf{x,y}\right)
e^{\frac{-i}{\hbar}\left(  \frac{\mathbf{P}\cdot\left(  \mathbf{x+y}\right)
}{2}-mt\right)  }\psi^{\beta\dagger\left(  h\right)  }\left(  \mathbf{x}%
\right)  \right. \tag{10}\\
&  \left.  -\psi^{\alpha\left(  h\right)  }\left(  \mathbf{x}\right)
\varphi_{\alpha\beta}^{\star}\left(  \mathbf{x,y}\right)  e^{\frac{-i}{\hbar
}\left(  \frac{\mathbf{P}\cdot\left(  \mathbf{x+y}\right)  }{2}-mt\right)
}\psi^{\beta\left(  e\right)  }\left(  \mathbf{x}\right)  \right]
d^{3}\mathbf{x}d^{3}\mathbf{y}\nonumber
\end{align}

\begin{quotation}
Remark 4: expression (10) is the theoretical basis of our model being
absolutely general and does not depends, in principle, on the hamiltonian
under consideration.
\end{quotation}

\section{Thermalization: new symmetries and emergent transformations}

To begin with and only in order to make a concise analysis of the construction
given before in [1], let us consider a Wannier exciton system. As is well
known, such system of excitons is characterized by the screening of the
crystal structure being well described by the following Schrodinger
equation$\left(  i\hbar\frac{\partial}{\partial t}-h\right)  \left\vert
\varphi\right\rangle =0$ that can be written using new displacement operators
defined as $\widetilde{D}\equiv D_{th}D_{\varphi}$ where $D_{th}$ is the
thermal unitary operator that arise as a consequence of the definition of the
squeezed-coherent states as transformed vacua under the automorphism group of
the commutation relations, as the thermofield dynamics case [5]. Then we have
\begin{equation}
\widetilde{D}^{\dag}\left(  i\hbar\frac{\partial}{\partial t}-h\right)
\widetilde{D}\left\vert 0\right\rangle =0\tag{11}%
\end{equation}
This transformation, shows specifically the group structure, namely, the
fundamental symmetries underlying the physics of the system. Clearly a
Bogoliubov-like transformation arises as in the pure excitonic case [1] or by
definition in the pure thermal case [5]:%
\begin{equation}
\binom{\widetilde{\psi}_{\alpha}^{\left(  e\right)  }}{\widetilde{\psi
}_{\alpha}^{\dagger\left(  h\right)  }}=\left(
\begin{array}
[c]{cc}%
\lambda\cos\varphi+\mu^{\ast}\frac{\varphi_{\alpha\beta}^{\ast}}{\varphi}%
\sin\varphi\text{ }e^{-\frac{i}{\hbar}\left(  \mathbf{P}\cdot\mathbf{x}%
-mt\right)  } & \mu\cos\varphi+\lambda^{\ast}\frac{\varphi_{\alpha\beta}%
^{\ast}}{\varphi}\sin\varphi\text{ }e^{-\frac{i}{\hbar}\left(  \mathbf{P}%
\cdot\mathbf{x}-mt\right)  }\\
-\lambda\frac{\varphi_{\alpha\beta}}{\varphi}\sin\varphi\text{ }e^{\frac
{i}{\hbar}\left(  \mathbf{P}\cdot\mathbf{x}-mt\right)  }+\mu^{\ast}\cos\varphi
& \lambda^{\ast}\cos\varphi-\mu\frac{\varphi_{\alpha\beta}}{\varphi}%
\sin\varphi\text{ }e^{\frac{i}{\hbar}\left(  \mathbf{P}\cdot\mathbf{x}%
-mt\right)  }%
\end{array}
\right)  \binom{\psi_{\alpha}^{\left(  e\right)  }}{\psi_{\alpha}%
^{\dagger\left(  h\right)  }}\tag{12}%
\end{equation}%
\[
\left\vert \lambda\right\vert ^{2}-\left\vert \mu\right\vert ^{2}=1
\]
The structure of above transformation is regulated by the same Green function
that defines the exciton wave function plus the coefficients $\lambda$ and
$\mu$ that are related with the eigenvalues of the exciton and fermion number
operators of the system $(N_{e},N_{h})$ given precisely the specific form of
the interaction hole-electron in the thermal case [6]. Introducing the
transformed fields via the displacement operator into the Schrodinger equation
we obtain schematically
\begin{equation}
\left[  \psi^{\dagger\left(  e\right)  }\widetilde{h}^{\left(  e\right)  }%
\psi^{\left(  e\right)  }+\psi^{\dagger\left(  h\right)  }\widetilde
{h}^{\left(  h\right)  }\psi^{\left(  h\right)  }+\psi^{\dagger\left(
e\right)  }Q\psi^{\dagger\left(  h\right)  }+\psi^{\left(  h\right)
}Q^{\dagger}\psi^{\left(  e\right)  }\right]  \left\vert 0\right\rangle
\tag{13}%
\end{equation}
where%
\[
\widetilde{h}_{\alpha\beta}^{\left(  e\right)  }\equiv m\delta_{\alpha\beta
}\left(  \left\vert \lambda\right\vert ^{2}\sin^{2}\varphi+\left\vert
\mu\right\vert ^{2}\cos^{2}\varphi\right)  -h_{\alpha\beta}\left(  \left\vert
\lambda\right\vert ^{2}+\left\vert \mu\right\vert ^{2}\right)  \cos2\varphi
\]
and%
\begin{equation}
Q_{\alpha\beta}^{th}\equiv-\left(  \left\vert \lambda\right\vert
^{2}+\left\vert \mu\right\vert ^{2}+\mu\lambda\right)  e^{\frac{i}{\hbar
}\left(  \mathbf{P}\cdot\mathbf{x}-mt\right)  }\left(  m\delta_{\alpha
}^{\ \gamma}-2h_{\alpha}^{\text{\ }\gamma}\right)  \frac{\sin\left(
2\varphi\right)  }{\varphi}\varphi_{\gamma\beta}\tag{14}%
\end{equation}
(and analogically for $\widetilde{h}_{\alpha\beta}^{\left(  h\right)  }$ and
$Q_{\alpha\beta}^{\dagger})$. We see that expression (14) must be zero if the
number of particles is conserved. Is not difficult to see that one condition
is $\frac{m}{2}=n_{f}h:$ the chemical potential $m$ is proportional to the
energy times the fermionic number of the system (the total energy considering
the binding). This is an equilibrium condition. The other one gives a
condition over the specific strength of the interaction electron-hole, namely:
$\sin\left(  2\varphi\right)  =0$ with $\varphi$ being the norm of the exciton
wave function defined by expression (5). Notice that both conditions are
independent of the thermal properties of the system. Other condition namely
$\left(  \left\vert \lambda\right\vert ^{2}+\left\vert \mu\right\vert ^{2}%
+\mu\lambda=0\right)  $ involves the thermal properties of the system [5]. And
this fact is far to be trivial due the behaviour of the transformations (12).
The concrete explanation of these conditions from the physical and
mathematical point of view will be part of a separated publications, and not
will be discussed here[6]. But the main points arising from expressions
(11-14) are:

i) transformations (14) control the general behaviour of the physical system,

ii) the group dependence of the transformation changes due to the basic wave
function of the exciton expression (5), that contains intrinsically the
electron hole interaction . Notice that this interaction is precisely the
building block of the Green function (4)and (6).

iii) facts i) and ii) reflect the conductance properties of \ the material
under consideration and the thermal influence.

From points i-iii) above, the model presented here can help to understand the
metal insulator transition in the thermal case. The transition from the
excitonic phase of the electron-hole system to the conducting situation .must
be characterized by the breaking of the pair , then, this fact is immediately
reflected in the changing of the transformations (21).We believe that this
effect is promising to be key to the interpretation and understanding of the
metal-insulator transition even in the thermal case[6].

\section{Acknowledgments}

We are very grateful to the organizers of ATOMS\ 2014 where part of this work
dedicated to Professor Julian Sereni was presented. Many thanks are also given
to Professor Illya G. Kaplan for important remarks in crucial part of this
work. We are very grateful to the Directorate of the JINR and BLTP\ for the
hospitality and financial support.

\section{References}

[1] Diego Julio Cirilo-Lombardo, Physics of Particles and Nuclei Letters,
2014, Vol. 11, No. 4, pp. 502--505 and references therein[1]

[2] There are several reviews on the subject, one of them is given by Massimo
Rontani and L. J. Sham in \textit{Novel Superfluids Volume 2}, edited by K. H.
Bennemann and J. B. Ketterson, International Series of Monographs on Physics
no. 157, Oxford University Press; A.A. Hight et al., Nature.(2012 Mar
21);483(7391):584-8. doi: 10.1038/nature10903.

[3] The preliminary ideas explaining the possibility of a concrete excitonic
condensation are in the following references: Keldysh, L. V. (1986). Contemp.
Phys., 27, 395; Keldysh, L. V. and Kopaev, Yu. V. (1964). Fiz. Tverd. Tela.,
6, 2791. (Sov.Phys. Solid State 6, 2219 (1965));. Keldysh, L. V. and Kozlov,
A. N. (1968). Zh. Eksp. i Teor. Fiz., 54, 978. (Sov.Phys.--JETP 27, 521
(1968)); Keldysh, L. V. and Kozlov, Z. N. (1967). Zh. Eksp. i Teor. Fiz.
Pisma, 5, 238.(Sov. Phys.--JETP Lett. 5, 190 (1968)).

[4] Klauder, J.R., Skagerstam, B.S.: Coherent States. World Scientific,
Singapore (1985); Klauder, J.R., Skagerstam, B.-S.K.: J. Phys. A: Math. Theor.
40, 2013 (2007); Klauder, J.R., Sudarshan, E.C.G.: Fundamentals of Quantum
Optics. Benjamin, New York (1968).

[5] See for example: G.W. Semenoff, H. Umezawa, Nucl.Phys. \textbf{B}220
(1983) 196-212 .

[6] Diego Julio Cirilo-Lombardo, in preparation.About This Shell

\end{document}